\documentclass[11pt,onecolumn]{IEEEtran}
\usepackage{amsmath,graphicx,amssymb,epsfig,setspace,multicol,cite,float,dsfont}
\usepackage[latin1]{inputenc}
\usepackage{graphicx}

\usepackage{schemabloc}  
\usepackage{epsfig}       
\usepackage{pst-all}      

\usepackage{tikz}
\usepackage{epstopdf}
\usetikzlibrary{positioning}

\DeclareTextSymbol{\degre}{T1}{6}
\DeclareTextSymbol{\degre}{OT1}{23}

\newcommand\tra{{\rm Tr}}

\newcommand\ve{{\rm vec}}
\newcommand\diag{{\rm Diag}}
\newcommand\e{{\rm E}}
\newcommand\pardef{ \stackrel{{\rm def}}{=} }

\newtheorem{lemma}{Lemma}

\def\QED{\mbox{\rule[0pt]{1.4ex}{1.4ex}}}

\usepackage{color}

\begin{document}

\title{Detailed proofs of paper \cite{Abeida2019}\\
Slepian-Bangs formula and Cram\'er Rao bound for circular and non-circular complex elliptical symmetric distributions
}

\author{Habti Abeida and Jean-Pierre Delmas}

\maketitle


\section{Useful relations and lemma}
\label{sec:Useful relations and lemma}
\subsection{Useful relations}
\label{sec:Useful relations}
We will make use of the following well known relations which hold for any conformable matrices ${\bf A}$,
${\bf B}$, ${\bf C}$ and ${\bf D}$.
\begin{equation}
\label{eq:ABC}
\ve({\bf A}{\bf B}{\bf C})=({\bf C}^T\otimes{\bf A})\ve({\bf B}),
\end{equation}
\begin{equation}
\label{eq:ABCD}
({\bf A}\otimes{\bf B})({\bf C}\otimes{\bf D})={\bf A}{\bf C} \otimes {\bf B}{\bf D},
\end{equation}
\begin{equation}
\label{eq:AB}
\tra({\bf A}{\bf B})=\ve^H({\bf A}^H)\ve({\bf B}),
\end{equation}
\begin{equation}
\label{eq:trABCD}
\tra({\bf A}{\bf B}{\bf C}{\bf D})=\ve^H({\bf A}^H)  ({\bf D}^T\otimes{\bf B}) \ve({\bf C}),
\end{equation}
\begin{equation}
\label{eq:A k B}
\tra({\bf A}\otimes{\bf B})=\tra({\bf A})\tra({\bf B}),
\end{equation}
\begin{equation}
\label{eq:tr K A  B}
\tra[{\bf K}({\bf A}\otimes{\bf B})]=\tra({\bf A}{\bf B}),
\end{equation}
where ${\bf K}$ is the vec-permutation matrix which transforms $\ve({\bf C})$ to $\ve({\bf C}^T)$
for any square matrix ${\bf C}$,

\begin{equation}
\label{eq:inv A B C D}
({\bf A}+{\bf B}{\bf C}{\bf D})^{-1}={\bf A}^{-1}-{\bf A}^{-1}{\bf B}({\bf C}^{-1}+{\bf D}{\bf A}^{-1}{\bf B})^{-1}{\bf D}{\bf A}^{-1},
\end{equation}
where ${\bf A}$, ${\bf C}$ and ${\bf C}^{-1}+{\bf D}{\bf A}^{-1}{\bf B}$ are assumed invertible.
%
\subsection{Useful lemma for the proof of Result 2}
\label{sec:Useful lemma for the proof of Result 2}
%
\begin{lemma}
\label{res:lemma1}
Let $\widetilde{\bf A}=\left(\begin{array}{cc}
{\bf A}_1&{\bf A}_2\\
{\bf A}_2^{*} &{\bf A}_1^{*} \\
\end{array}\right)$
and $\widetilde{\bf B}=\left(\begin{array}{cc}
{\bf B}_1&{\bf B}_2\\
{\bf B}_2^{*} &{\bf B}_1^{*} \\
\end{array}\right)$
be two $2M\times2M$ partitioned matrices with ${\bf A}_1$ and ${\bf B}_1$ are $M\times M$ Hermitian matrices,
${\bf A}_2$ and ${\bf B}_2$ are $M\times M$ complex symmetric matrices,
and suppose that
${\bf y}\sim \mathbb{C}\mathcal{N}_{M}({\bf 0},{\bf I})$.
Then
\begin{equation}
\label{eq:EyayyBy}
\e[(\widetilde{\bf y}^{H}\widetilde{\bf A}\widetilde{\bf y})(\widetilde{\bf y}^H\widetilde{\bf B}\widetilde{\bf y})]
=\tra(\widetilde{\bf A})\tra(\widetilde{\bf B})+2\tra(\widetilde{\bf A}\widetilde{\bf B}),
\end{equation}
where
$\widetilde{\bf y}\pardef ({\bf y}^T,{\bf y}^H)^T$.
\end{lemma}
\emph{Proof:}

We get from \eqref{eq:trABCD} then \eqref{eq:ABCD}
\begin{equation}
\label{eq:EyayyBy suite}
\e[(\widetilde{\bf y}^{H}\widetilde{\bf A}\widetilde{\bf y})(\widetilde{\bf y}^H\widetilde{\bf B}\widetilde{\bf y})]
=\tra[(\widetilde{\bf A}^T\otimes\widetilde{\bf B})
\e(\widetilde{\bf y}^*\widetilde{\bf y}^T\otimes\widetilde{\bf y}\widetilde{\bf y}^H
)],
\end{equation}
where from e.g. \cite[Appendix B]{Abeida2006}
\begin{equation}
\label{eq:property y}
\e(\widetilde{\bf y}^*\widetilde{\bf y}^T\otimes\widetilde{\bf y}\widetilde{\bf y}^H)
=
{\bf I}\otimes {\bf I}+{\bf K}({\bf J}'\otimes{\bf J}')({\bf I}\otimes {\bf I})
+\ve({\bf I})\ve^T({\bf I}),
\end{equation}
where
${\bf J}'\pardef\left(\begin{array}{cc}
{\bf 0}&{\bf I}\\
{\bf I}&{\bf 0} \\
\end{array}\right)$.
Plugging \eqref{eq:property y} in \eqref{eq:EyayyBy suite}, we get:
\begin{eqnarray}
\nonumber
\e[(\widetilde{\bf y}^{H}\widetilde{\bf A}\widetilde{\bf y})(\widetilde{\bf y}^H\widetilde{\bf B}\widetilde{\bf y})]
&=&
\tra[(\widetilde{\bf A}^T\otimes\widetilde{\bf B})
({\bf I}\otimes {\bf I})]
+
\tra[(\widetilde{\bf A}^T\otimes\widetilde{\bf B}){\bf K}({\bf J}'\otimes{\bf J}')({\bf I}\otimes {\bf I})]
\\
\label{eq:EyayyBy y bis}
&+&
\tra[(\widetilde{\bf A}^T\otimes\widetilde{\bf B})\ve({\bf I})\ve^T({\bf I})],
\end{eqnarray}
where we have successively
\[
\tra[(\widetilde{\bf A}^T\otimes\widetilde{\bf B})
({\bf I}\otimes {\bf I})]
=
\tra(\widetilde{\bf A})\tra(\widetilde{\bf B})
\]
from \eqref{eq:ABCD} and \eqref{eq:A k B},
\[
\tra[(\widetilde{\bf A}^T\otimes\widetilde{\bf B}){\bf K}({\bf J}'\otimes{\bf J}')({\bf I}\otimes {\bf I})]
=\tra(\widetilde{\bf A}\widetilde{\bf B})
\]
from \eqref{eq:ABCD}, \eqref{eq:tr K A  B} and ${\bf J}'\widetilde{\bf A}^T{\bf J}'=\widetilde{\bf A}$,
and
\[
\tra[(\widetilde{\bf A}^T\otimes\widetilde{\bf B})\ve({\bf I})\ve^T({\bf I})]
=\tra(\widetilde{\bf A}\widetilde{\bf B})
\]
from \eqref{eq:trABCD}.
Plugging these three expressions in \eqref{eq:EyayyBy y bis}, \eqref{eq:EyayyBy} follows.
\hfill
\QED
%
\section{Proof of Result 1 and Eq. (5) of \cite{Abeida2019}}
\label{sec:Proof of Result 1}
Since a linear transform in $\mathbb{R}^{2M}$ is tantamount to $\mathbb{R}$-linear transform in $\mathbb{C}^{M}$, the definition of
GCES given in \cite{Ollila2004} is equivalent to saying that\footnote{Note that if ${\bf \Phi}={\bf 0}$, ${\bf z}$
is C-CES distributed.}
\begin{equation}
\label{eq:stochastic representation proof}
{\bf z}={\boldsymbol \mu}+{\bf \Psi}{\bf z}_0 + {\bf \Phi}{\bf z}_0^*,
\end{equation}
where ${\bf \Psi}$ and ${\bf \Phi}$ are $M\times M$ fixed complex-valued matrices and ${\bf z}_0$ is a complex spherical distributed r.v. with stochastic representation ${\bf z}_0=_{d}\mathcal{R}{\bf u}$ \cite[th. 3]{Ollila2012}.
Since $\e({\bf u}{\bf u}^H)=\frac{1}{M}{\bf I}$ and $\e({\bf u}{\bf u}^T)={\bf 0}$ \cite[lemma 1b]{Ollila2012}, we get if
$\e(\mathcal{R}^2) < \infty$,
\begin{equation}
\label{eq:Sigma Omega}
{\bf \Sigma}
={\bf A}{\bf A}^H
=\frac{\e(\mathcal{R}^2)}{N\sigma_c}
\left({\bf \Psi}{\bf \Psi}^H+{\bf \Phi}{\bf \Phi}^H\right)
\ \ \mbox{and} \ \
{\bf \Omega}
=
{\bf A}{\bf \Delta}_{\kappa}{\bf A}^T
=
\frac{\e(\mathcal{R}^2)}{N\sigma_c}
\left({\bf \Psi}{\bf \Phi}^T+{\bf \Psi}{\bf \Phi}^T\right),
\end{equation}
where $\sigma_c$ is defined by  $\e[({\bf z}\!-\!{\boldsymbol \mu})({\bf z}\!-\!{\boldsymbol \mu})^H]=\sigma_c {\bf \Sigma}$
and $\e[({\bf z}\!-\!{\boldsymbol \mu})({\bf z}\!-\!{\boldsymbol \mu})^T]=\sigma_c {\bf \Omega}$ whose value is
$\e(\mathcal{R}^2)/N$ \cite[(14)]{Ollila2012}. Consequently \eqref{eq:Sigma Omega} reduces to
\begin{equation}
\label{eq:phi et psi}
{\bf A}{\bf A}^H
=
{\bf \Psi}{\bf \Psi}^H+{\bf \Phi}{\bf \Phi}^H
\ \ \mbox{and} \ \
{\bf A}{\bf \Delta}_{\kappa}{\bf A}^T
=
{\bf \Psi}{\bf \Phi}^T+{\bf \Psi}{\bf \Phi}^T.
\end{equation}
By the one to one change of variable (because ${\bf A}$ is nonsingular):
${\bf \Psi}'={\bf A}{\bf \Psi}$ and ${\bf \Phi}'={\bf A}{\bf \Phi}$, \eqref{eq:phi et psi} is equivalent to:
\begin{equation}
\label{eq:phi et psi equivalent}
{\bf I}
=
{\bf \Psi}'{\bf \Psi}^{'H}+{\bf \Phi}{\bf \Phi}^{'H}
\ \ \mbox{and} \ \
{\bf \Delta}_{\kappa}
=
{\bf \Psi}'{\bf \Phi}^{'T}+{\bf \Psi}'{\bf \Phi}^{'T}.
\end{equation}
It is clear that the solution of \eqref{eq:phi et psi equivalent} is not unique,
but we can look for solutions in real-valued diagonal form
$({\bf \Psi},{\bf \Phi})=({\bf \Delta}_1,{\bf \Delta}_2)$ with
\begin{equation}
\label{eq:phi et psi equivalent diag}
{\bf I}
=
{\bf \Delta}_1^2+{\bf \Delta}_2^2
\ \ \mbox{and} \ \
{\bf \Delta}_{\kappa}
=
2{\bf \Delta}_1{\bf \Delta}_2,
\end{equation}
whose solutions are
${\bf \Delta}_1=\frac{{\bf \Delta}_{+}+{\bf \Delta}_{-}}{2}$
and ${\bf \Delta}_2=\frac{{\bf \Delta}_{+}-{\bf \Delta}_{-}}{2}$
where ${\bf \Delta}_{+} \pardef \sqrt{{\bf I}+{\bf \Delta}_{\kappa}}$
and ${\bf \Delta}_{-} \pardef \sqrt{{\bf I}-{\bf \Delta}_{\kappa}}$.
Consequently
\begin{equation}
\label{eq:SR theorem}
{\bf z}=_d{\boldsymbol \mu}+\mathcal{R}[{\bf \Psi}{\bf u}+{\bf \Phi}{\bf u}^*]
={\boldsymbol \mu}+\mathcal{R}{\bf A}[{\bf \Delta}_1{\bf u}+{\bf \Delta}_2{\bf u}^*].
\end{equation}
If $\e(\mathcal{R}^2)$ is not finite,
the scatter and pseudo-scatter matrices of ${\bf z}$ given  by \eqref{eq:SR theorem} are also
${\bf \Sigma}
={\bf A}{\bf A}^H$ and
${\bf \Omega}
=
{\bf A}{\bf \Delta}_{\kappa}{\bf A}^T$, respectively.
\hfill
\QED

From the eigenvalue decomposition
$\left(\begin{array}{cc}
\!{\bf I}\!&\! {\bf \Delta}_{\kappa}\!\\
\!{\bf \Delta}_{\kappa}\!&\!  {\bf I} \!\\
\end{array}
\right)
\!=\!
\left[
\frac{1}{\sqrt{2}}
\left(\begin{array}{cc}
\!{\bf I}\!&\! {\bf I}\!\\
\!{\bf I}\!&\!  -{\bf I} \!\\
\end{array}
\!\right)
\!
\right]
\!
\left(\begin{array}{cc}
\!{\bf I}+{\bf \Delta}_{\kappa}\!&\! {\bf 0}\!\\
\!{\bf 0}\!&\!  {\bf I}+{\bf \Delta}_{\kappa} \!\\
\end{array}
\right)
\!\left[
\!
\frac{1}{\sqrt{2}}
\left(\begin{array}{cc}
\!{\bf I}\!&\! {\bf I}\!\\
\!{\bf I}\!&\!  -{\bf I} \!\\
\end{array}
\!\right)\!
\right]$, we deduce from
$\widetilde{\bf \Gamma}
= \left(\begin{array}{cc}
\!{\bf A}\!&\! {\bf 0}\!\\
\!{\bf 0}\!&\!  {\bf A}^{*} \!\\
\end{array}
\right)
\!\!
\left(\begin{array}{cc}
\!{\bf I}\!&\! {\bf \Delta}_{\kappa}\!\\
\!{\bf \Delta}_{\kappa}\!&\!  {\bf I} \!\\
\end{array}
\right)
\!\!
\left(\begin{array}{cc}
\!{\bf A}^H\!&\! {\bf 0}\!\\
\!{\bf 0}\!&\!  {\bf A}^{T} \!\\
\end{array}
\right)$
that
$\widetilde{\bf \Gamma}^{1/2}
=
\left(\begin{array}{cc}
\!{\bf A}\!&\! {\bf 0}\!\\
\!{\bf 0}\!&\!  {\bf A}^{*} \!\\
\end{array}
\right)
\!
\left(\begin{array}{cc}
\!\!{\bf \Delta}_1\!&\! {\bf \Delta}_2\!\!\\
\!\!{\bf \Delta}_2\!&\!  {\bf \Delta}_1 \!\!\\
\end{array}
\right)$.
Consequently, the stochastic representation
 ${\bf z}=_{d}{\boldsymbol \mu}+\mathcal{R}{\bf A}{\bf v}$
is equivalent to
\begin{equation}
\label{eq:SR}
\widetilde{\bf z}=_{d}\widetilde{\boldsymbol \mu}+\mathcal{R}\widetilde{\bf \Gamma}^{1/2}\widetilde{\bf u}
\end{equation}
with $\widetilde{\bf u}\pardef({\bf u}^T,{\bf u}^H)^T$.
It follows directly
$\frac{1}{2}(\tilde{\bf z}-\tilde{\bf \boldsymbol \mu})^H\tilde{\bf \Gamma}^{-1}(\tilde{\bf z}-\tilde{\bf \boldsymbol \mu})
=_{d}\frac{1}{2}\mathcal{R}^2\|\widetilde{\bf u}\|^2 =\mathcal{Q}$.
\hfill
\QED
%
\section{Proof of Result 2}
\label{sec:Proof of Result 2}
%
To prove this result, we follows the different steps of \cite[sec. 3]{Besson2013}.
First, we cheek that the p.d.f.  $p({\bf z};{\bf \boldsymbol{\alpha}})$ satisfies the ''regularity'' condition
\begin{equation}
\label{eq:regcondit}
\e\left(\frac{\partial \log p({\bf z};{\bf \boldsymbol{\alpha}})}{\partial \alpha_k}\right)=0.
\end{equation}
Taking the derivative of the p.d.f. \cite[(1)]{Abeida2019} w.r.t.  $\alpha_k$,  yields
\begin{equation}
\label{eq:firstderi1}
\frac{\partial \log p({\bf z};{\bf \boldsymbol{\alpha}})}{\partial \alpha_k}= -\frac{1}{2}\tra(\widetilde{\bf \Gamma}^{-1}\widetilde{\bf \Gamma}_{k})+\phi(\tilde{\eta})\frac{\partial \tilde{\eta}}{\partial \alpha_k}.
\end{equation}
It follows from the definition of $\tilde{\eta}$ that
\begin{equation}
\label{eq:Derivaeta}
\frac{\partial \tilde{\eta}}{\partial \alpha_k}
=-{\rm Re}\left(\tilde{\boldsymbol \mu}_k^H
 {\widetilde{\bf \Gamma}}^{-1}(\tilde{\bf z}-\tilde{\bf \boldsymbol{\mu}})  \right)
-\frac{1}{2}(\tilde{\bf z}-\tilde{\bf \boldsymbol{\mu}})^{H}{\widetilde{\bf \Gamma}}^{-1}{\widetilde{\bf \Gamma}}_k{\widetilde{\bf \Gamma}}^{-1} (\tilde{\bf z}-\tilde{\bf \boldsymbol{\mu}}),
\end{equation}
where
$\widetilde{\boldsymbol \mu}_k
\pardef
\frac{\partial \widetilde{\boldsymbol \mu}}{\partial \alpha_k}$
and
$\widetilde{\boldsymbol \Gamma}_k
\pardef
\frac{\partial \widetilde{\boldsymbol \Gamma}}{\partial \alpha_k}$.
Making use of the extended stochastic representation \eqref{eq:SR}, the second term of \eqref{eq:Derivaeta} is given by
\begin{equation}
\label{eq:eta bis}
\frac{1}{2}(\tilde{\bf z}-\tilde{\bf \boldsymbol{\mu}})^{H}{\widetilde{\bf \Gamma}}^{-1}{\widetilde{\bf \Gamma}}_k{\widetilde{\bf \Gamma}}^{-1} (\tilde{\bf z}-\tilde{\bf \boldsymbol{\mu}})
=_d\frac{1}{2}\mathcal{Q}\tilde{\bf u}^H\widetilde{\bf H}_k\tilde{\bf u}
\end{equation}
where
$\widetilde{\bf H}_k
\pardef
{\widetilde{\bf \Gamma}}^{-1/2}{\widetilde{\bf \Gamma}}_k{\widetilde{\bf \Gamma}}^{-1/2}$.
Thus using $\tilde{\eta}=_d \mathcal{Q}$ \cite[(5)]{Abeida2019}, we get:
\begin{equation}
\label{eq:phi eta}
\e\left(\phi(\tilde{\eta})\frac{\partial \tilde{\eta}}{\partial \alpha_k}\right)
=-\e\left(\mathcal{Q}^{1/2}\phi(\mathcal{Q}){\rm Re}(\tilde{\boldsymbol \mu}_k^H\widetilde{\bf \Gamma}^{-1/2}\tilde{\bf u})\right)-
\frac{1}{2}\e[\mathcal{Q}\phi(\mathcal{Q})\tilde{\bf u}^H\widetilde{\bf H}_k\tilde{\bf u}].
\end{equation}
Since $\mathcal{Q}$ and  ${\bf u}$ are independent, $\mathcal{Q}$ and $\tilde{\bf u}$ are also independent. It follows then from $\e(\tilde{\bf u})={\bf 0}$, $\e(\tilde{\bf u}\tilde{\bf u}^H)=\frac{1}{M}{\bf I}$ and $\e(\mathcal{Q}\phi(\mathcal{Q}))=-M$ \cite[(11)]{Besson2013} that
\[
\e\left(\mathcal{Q}^{1/2}\phi(\mathcal{Q}){\rm Re}(\tilde{\boldsymbol \mu}_k^H
\widetilde{\bf \Gamma}^{-1/2}\tilde{\bf u})\right)=0
\]
and
\[
\e[\mathcal{Q}\phi(\mathcal{Q})\tilde{\bf u}^H\widetilde{\bf H}_k\tilde{\bf u}]
=
\e[\mathcal{Q}\phi(\mathcal{Q})]
\tra[\widetilde{\bf H}_k\e(\tilde{\bf u}\tilde{\bf u}^H)]
=-\tra(\widetilde{\bf H}_k)
=-\tra(\widetilde{\bf \Gamma}^{-1}\widetilde{\bf \Gamma}_k).
\]
Thus
\begin{equation}
\label{eq:Rela1}
\e\left(\phi(\tilde{\eta})\frac{\partial \tilde{\eta}}{\partial \alpha_k}\right)
=\frac{1}{2}\tra(\widetilde{\bf \Gamma}^{-1}\widetilde{\bf \Gamma}_k),
\end{equation}
which proves \eqref{eq:regcondit}.

Now, we evaluate the elements of the FIM. It follows from \eqref{eq:firstderi1}, using \eqref{eq:Rela1},  that
\begin{equation}
\label{eq:FIMDerexp}
[{\bf I}^{\rm NC}_{\rm CES}]_{k,l}
=\e\left(\frac{\partial \log p({\bf z};{\bf \boldsymbol{\alpha}})}{\partial \alpha_k}
\frac{\partial \log p({\bf z};{\bf \boldsymbol{\alpha}})}{\partial \alpha_l}\right)
=-\frac{1}{4}\tra(\widetilde{\bf \Gamma}^{-1}\widetilde{\bf \Gamma}_k)\tra(\widetilde{\bf \Gamma}^{-1}\widetilde{\bf \Gamma}_l)+\e\left(\phi^2(\tilde{\eta})\frac{\partial \tilde{\eta}}{\partial \alpha_k}\frac{\partial \tilde{\eta}}{\partial \alpha_l}\right).
\end{equation}
It follows from \eqref{eq:SR} that
$\widetilde{\bf \Gamma}^{-1/2}(\widetilde{\bf z}-\widetilde{\boldsymbol \mu})=_d
\sqrt{\mathcal{Q}}\ \widetilde{\bf u}$ and hence from \eqref{eq:Derivaeta} we get
\begin{eqnarray}
\nonumber
\phi^2(\tilde{\eta})\frac{\partial \tilde{\eta}}{\partial \alpha_k}\frac{\partial \tilde{\eta}}{\partial \alpha_l}
&=_d&
\mathcal{Q}\phi^2(\mathcal{Q}) {\rm Re}\left(\tilde{\boldsymbol \mu}_k^H \widetilde{\bf \Gamma}^{-1/2}\tilde{\bf u} \right)
{\rm Re}\left(\tilde{\boldsymbol \mu}_l^H \widetilde{\bf \Gamma}^{-1/2}\tilde{\bf u} \right)
\\
\nonumber
&+&
\frac{1}{2}\mathcal{Q}^{3/2}\phi^2(\mathcal{Q}) {\rm Re}\left(\tilde{\boldsymbol \mu}_l^H\widetilde{\bf \Gamma}^{-1/2}
\tilde{\bf u} \right)[\tilde{\bf u}^H\widetilde{\bf H}_k
\tilde{\bf u}]
+
\frac{1}{2}\mathcal{Q}^{3/2}\phi^2(\mathcal{Q}) {\rm Re}\left(\tilde{\boldsymbol \mu}_k^H\widetilde{\bf \Gamma}^{-1/2}
\tilde{\bf u}\right)[\tilde{\bf u}^H\widetilde{\bf H}_l\tilde{\bf u}]
\\
\label{eq:TermCRB}
&+&
\frac{1}{4}\mathcal{Q}^2\phi^2(\mathcal{Q})[\tilde{\bf u}^H\widetilde{\bf H}_k\tilde{\bf u}][\tilde{\bf u}^H\widetilde{\bf H}_l\tilde{\bf u}].
\end{eqnarray}
The first term of \eqref{eq:TermCRB} can be further simplified as
\[
{\rm Re}\left(\tilde{\boldsymbol \mu}_k^H \widetilde{\bf \Gamma}^{-1/2}\tilde{\bf u} \right)
{\rm Re}\left(\tilde{\boldsymbol \mu}_l^H\widetilde{\bf \Gamma}^{-1/2}\tilde{\bf u} \right)
=
\frac{1}{2}{\rm Re}\left(\tilde{\boldsymbol \mu}_k^H \widetilde{\bf \Gamma}^{-1/2}\tilde{\bf u}\tilde{\bf u}^H
\widetilde{\bf \Gamma}^{-1/2}\tilde{\boldsymbol \mu}_l\right)
+
\frac{1}{2}{\rm Re}\left(\tilde{\boldsymbol \mu}_k^T\widetilde{\bf \Gamma}^{-*1/2}\tilde{\bf u}^*\tilde{\bf u}^H
\widetilde{\bf \Gamma}^{-1/2}\tilde{\boldsymbol \mu}_l\right),
\]
and thanks to the independence between $\mathcal{Q}$ and $\tilde{\bf u}$, the expected value of the first term of
\eqref{eq:TermCRB} is given by
\begin{eqnarray}
\nonumber
\e[\mathcal{Q}\phi^2(\mathcal{Q})]
\e
\left(
{\rm Re}\left(\tilde{\boldsymbol \mu}_k^H \widetilde{\bf \Gamma}^{-1/2}\tilde{\bf u} \right)
{\rm Re}\left(\tilde{\boldsymbol \mu}_l^H \widetilde{\bf \Gamma}^{-1/2}\tilde{\bf u} \right)
\right)
&=&
\\
\label{eq:FIMDerexp first term}
&&
\hspace{-8cm}
\frac{\e[\mathcal{Q}\phi^2(\mathcal{Q})]}{2M}{\rm Re}\left(\tilde{\boldsymbol \mu}_k^H{\widetilde{\bf \Gamma}}^{-1}\tilde{\boldsymbol \mu}_l\right)
+
\frac{\e[\mathcal{Q}\phi^2(\mathcal{Q})]}{2M}
{\rm Re}\left(\tilde{\boldsymbol \mu}_k^T{\widetilde{\bf \Gamma}}^{-*}{\bf J}'\tilde{\boldsymbol \mu}_l\right)
=
\frac{\e[\mathcal{Q}\phi^2(\mathcal{Q})]}{M}{\rm Re}\left(\tilde{\boldsymbol \mu}_k^H{\widetilde{\bf \Gamma}}^{-1}\tilde{\boldsymbol \mu}_l\right),
\end{eqnarray}
using $\e(\tilde{\bf u}\tilde{\bf u}^H)
=\frac{1}{M}{\bf I}$
and $\e(\tilde{\bf u}^*\tilde{\bf u}^H)
=\frac{1}{M}{\bf J}$,
$\widetilde{\bf \Gamma}^{-*1/2}{\bf J}'\widetilde{\bf \Gamma}^{-1/2}={\widetilde{\bf \Gamma}}^{-*}{\bf J}'$
and ${\bf J}'\tilde{\boldsymbol \mu}_l
=\tilde{\boldsymbol \mu}_l^*$.
The expected value of the second and third terms of \eqref{eq:TermCRB} are zero because the third-order moments of ${\bf u}$ are zero.
Because
${\bf y}=_d\|{\bf y}\| {\bf u}$, where $\|{\bf y}\|$ and ${\bf u}$ are independent when ${\bf y}\sim \mathbb{C}\mathcal{N}_{M}({\bf 0},{\bf I})$, we get
\[
\e[(\tilde{\bf u}^{H}\widetilde{\bf H}_k\tilde{\bf u})
(\tilde{\bf u}^H\widetilde{\bf H}_l\tilde{\bf u})]
=
\frac{1}{\e(\|{\bf y}\|^4)}\e[(\widetilde{\bf y}^{H}\widetilde{\bf H}_k\widetilde{\bf y})
(\widetilde{\bf y}^H\widetilde{\bf H}_l\widetilde{\bf y})].
\]
Noting that $\widetilde{\bf H}_k$ and $\widetilde{\bf H}_l$ are structured as
$\widetilde{\bf A}$ and $\widetilde{\bf B}$ of the Lemma \ref{res:lemma1}, this lemma applies to the couples
$(\widetilde{\bf H}_k,\widetilde{\bf H}_l)$ and $({\bf I},{\bf I}$) giving
$\e[(\widetilde{\bf y}^{H}\widetilde{\bf H}_k\widetilde{\bf y})
(\widetilde{\bf y}^H\widetilde{\bf H}_l\widetilde{\bf y})]=\tra(\widetilde{\bf H}_k)\tra(\widetilde{\bf H}_l)
+2\tra(\widetilde{\bf H}_k\widetilde{\bf H}_l)$
and $\e[\|\widetilde{\bf y}\|^4]=4M(M+1)$.
Consequently the expected value of the last term of \eqref{eq:TermCRB} is given by
\begin{eqnarray}
\nonumber
\e\left(\frac{1}{4}\mathcal{Q}^2\phi^2(\mathcal{Q})[\tilde{\bf u}^H\widetilde{\bf H}_k\tilde{\bf u}][\tilde{\bf u}^H\widetilde{\bf H}_l\tilde{\bf u}]\right)
&=&
\frac{\e(\mathcal{Q}^2\phi^2(\mathcal{Q}))}{4M(M+1)}
\left(\tra(\widetilde{\bf H}_k)\tra(\widetilde{\bf H}_l)+2\tra(\widetilde{\bf H}_k\widetilde{\bf H}_l)\right)
\\
\label{eq:FIMDerexp last term}
&=&
\frac{\e(\mathcal{Q}^2\phi^2(\mathcal{Q}))}{4M(M+1)}
\left(\tra(\widetilde{\bf \Gamma}_k\widetilde{\bf \Gamma}^{-1})\tra(\widetilde{\bf \Gamma}_l\widetilde{\bf \Gamma}^{-1})+2\tra(\widetilde{\bf \Gamma}_k\widetilde{\bf \Gamma}^{-1}\widetilde{\bf \Gamma}_l\widetilde{\bf \Gamma}^{-1})\right).
\end{eqnarray}
Gathering \eqref{eq:FIMDerexp first term} \eqref{eq:FIMDerexp last term} in \eqref{eq:FIMDerexp} concludes the proof.
\hfill
\QED
%
\section{Proof of Eq. (9) of \cite{Abeida2019}}
\label{sec:Proof of Eq. (9)}
%
Using that \cite[(4)]{Abeida2019} is a p.d.f. with $\int_{0}^{\infty}\delta^{-1}_{M,g}\mathcal{Q}^{M-1}_tg(\mathcal{Q}_t)d\mathcal{Q}_t=1$ and that $\e(\mathcal{Q})=\e(\mathcal{R}^2)< \infty$, we get
\begin{equation}
\label{eq:expectation}
\e(\mathcal{Q}\phi(\mathcal{Q}))
=\int_{0}^{\infty}\delta^{-1}_{M,g}\mathcal{Q}^{M}g'(\mathcal{Q})d\mathcal{Q}
=\left[\delta^{-1}_{M,g}\mathcal{Q}^{M}g(\mathcal{Q})\right]_{0}^{\infty}
-
M\int_{0}^{\infty}\delta^{-1}_{M,g}\mathcal{Q}^{M-1}g(\mathcal{Q})d\mathcal{Q}
=-M.
\end{equation}
It follows from Cauchy-Schwarz inequality that
\begin{equation}
\label{eq:Cauchy-Schwarz inequality}
M^2=(\e(\mathcal{Q}\phi(\mathcal{Q})))^2\leq\e(\mathcal{Q})\e(\mathcal{Q}\phi^2(\mathcal{Q}))=\e(\mathcal{Q})M\xi_1.
\end{equation}
Next, note that
\begin{equation}
\label{eq:relation}
\e(\mathcal{Q})=\int_{0}^{\infty}\delta^{-1}_{M,g}\mathcal{Q}^{M}g(\mathcal{Q})d\mathcal{Q}
=\delta^{-1}_{M,g}\delta_{M+1,g}\int_{0}^{\infty}\delta^{-1}_{M+1,g}\mathcal{Q}^{M}g(\mathcal{Q})d\mathcal{Q}
=\delta^{-1}_{M,g}\delta_{M+1,g}=M.
\end{equation}
Plugging \eqref{eq:relation} in \eqref{eq:Cauchy-Schwarz inequality} proves Eq. (9) of \cite{Abeida2019}.
\hfill
\QED

\section{Proof of Result 4}
\label{sec:Proof of Result 4}
%
Because $\xi_2=1$ for Gaussian distributions, we get for NC-CES distributions:
\begin{equation}
\label{eq:I NC CES}
{\bf I}^{\rm NC}_{\rm CES}({\boldsymbol \alpha_2})-{\bf I}^{\rm NC}_{\rm CN}({\boldsymbol \alpha_2})
=\frac{\xi_2-1}{2}
\left(\frac{d \ve(\widetilde{\bf \Gamma})}{d {\boldsymbol \alpha}_2^T}\right)^H
\left(
({\widetilde{\bf \Gamma}}^{-T}\otimes\widetilde{\bf \Gamma}^{-1})
+\frac{1}{2}\ve(\widetilde{\bf \Gamma}^{-1})\ve^H(\widetilde{\bf \Gamma}^{-1})
\right)
\frac{d \ve(\widetilde{\bf \Gamma})}{d {\boldsymbol \alpha}_2^T}
\end{equation}
where
$({\widetilde{\bf \Gamma}}^{-T}\otimes\widetilde{\bf \Gamma}^{-1})
+\frac{1}{2}\ve(\widetilde{\bf \Gamma}^{-1})\ve^H(\widetilde{\bf \Gamma}^{-1})$ is positive definite.
Replacing ${\widetilde{\bf \Gamma}}$ by ${\bf \Gamma}$, the proof is identical  for C-CES distributions.
\hfill
\QED
%
\section{Proof of Result 5}
\label{sec:Proof of Result 5}
We note first that the general  expressions of the SCRB  proved here is valid for arbitrary parameterization of 
${\bf A}_{\theta}$ if the  real-valued parameter of interest $\boldsymbol{\theta}\in \mathbb{R}^L$ is characterized by the subspace generated by the columns of the full column rank $M \times K$ matrix
${\bf A}_{\theta}$ with $K<M$.
It can be applied for example to near or far-field DOA modeling  with scalar or vector-sensors for an arbitrary
number of parameters per source $s_{t,k}$ (with ${\bf s}_t \pardef (s_{t,1},..,s_{t,K})^T$ and many other modelings as the SIMO and MIMO modelings. Let us start with the  circular case for which  ${\bf \Omega}={\bf 0}$  and thus
$\widetilde{\bf \Gamma}
=
\diag({\bf \Sigma},{\bf \Sigma}^*)$ where ${\bf \Sigma}={\bf A}_{\theta}{\bf R}_s{\bf A}^H_{\theta}+\sigma^2_n{\bf I}$. The SCRB    form for this case can be then written   through the  compact expression of the general FIM  given in Result 2, using \eqref{eq:ABC} and \eqref{eq:ABCD}, as follows:
\begin{equation}
\label{eq:GenSCRBCCES}
\frac{1}{T}{\rm SCRB}^{-1}_{\rm CES}({\boldsymbol \alpha})
=
\left(
\frac{d \ve({\bf \Sigma})}{d{\boldsymbol \alpha}^T}\right)^H
\left(\xi_2({{\bf \Sigma}}^{-T}\otimes{\bf \Sigma}^{-1})+
(\xi_{2}\!-\!1)\ve({\bf \Sigma}^{-1}\!)\ve^H({\bf \Sigma}^{-1}\!)
\right)\left(\frac{d \ve({\bf \Sigma})}{d {\boldsymbol \alpha}^T}
\!\!\right).
\end{equation}
The SCRB  of ${\boldsymbol \theta}$ alone can be deduced from  \eqref{eq:GenSCRBCCES} as follows:
\begin{equation}
\label{eq:SCRBCCES}
\frac{1}{T}{\rm SCRB}^{-1}_{\rm CES}({\boldsymbol \theta})=
{\bf G}^{H}{\boldsymbol \Pi}^{\bot}_{{\bf \Delta}}{\bf G},
\end{equation}
with ${\bf G}\pardef{\bf T}^{1/2}_i({{\bf \Sigma}}^{-T/2}\otimes{\bf \Sigma}^{-1/2})\frac{\partial \ve({\bf \Sigma})}{\partial {\bf \boldsymbol{\theta}}^T}$ and
${\bf \Delta}\pardef{\bf T}^{1/2}_i({{\bf \Sigma}}^{-T/2}\otimes{\bf \Sigma}^{-1/2})\frac{\partial \ve({\bf \Sigma})}{\partial {\bf \boldsymbol{\alpha}}^T_n}$ where
\begin{equation}
\label{eq:def Ti}
{\bf T}_i\pardef\xi_{2} {\bf I}+(\xi_{2}-1)\ve({\bf I})\ve^T({\bf I}).
\end{equation}
Let's further partition
the matrix ${\bf \Delta}$ as  ${\bf \Delta}={\bf T}^{1/2}_i({\bf \Sigma}^{-T/2}\otimes{\bf \Sigma}^{-1/2}) \left[
\frac{\partial \ve({\bf \Sigma})}{\partial \boldsymbol{\rho}^T} \mid
\frac{\partial \ve({\bf \Sigma})}{\partial \sigma_n^2}
\right]
\pardef \left[{\bf V} \mid {\bf u}_n\right]$.   In the sequel, the proofs presented here follow the lines of the proof  presented in  \cite{Stoica2001} for circular Gaussian distributed observations.  It follows from \cite[rel. (14)]{Stoica2001}
 that
 \begin{equation}
\label{eq:ProjectorExt}
 {\bf \Pi}^{\bot}_{{\bf \Delta}}={\bf \Pi}^{\bot}_{{\bf V}}-\frac{{\bf \Pi}^{\bot}_{{\bf V}}{\bf u}_n{\bf u}^H_n{\bf \Pi}^{\bot}_{{\bf V}}}{{\bf u}^H_n{\bf \Pi}^{\bot}_{{\bf V}}{\bf u}_n}.
\end{equation}
Using $\frac{\partial \ve({\bf \Sigma})}{\partial \sigma_n^2}=\ve({\bf I})$, we obtain
\begin{equation}
\label{eq:vecun}
{\bf u}_n={\bf T}^{1/2}_i\ve({\bf \Sigma}^{-1}).
\end{equation}
Consequently using \eqref{eq:SCRBCCES} and \eqref{eq:ProjectorExt}, if ${\bf g}_k$ denotes the \emph{kth} column of ${\bf G}$, the $(k,l)$ element of ${\rm SCRB}^{-1}_{\rm CES}({\boldsymbol \alpha})$ can be written elementwise as
\begin{equation}
\label{eq:CRBGen}
\frac{1}{T}\left[{\rm SCRB}^{-1}_{\rm CES}({\boldsymbol \theta})\right]_{k,l}
={\bf g}_k^H{\bf \Pi}^{\bot}_{{\bf V}}{\bf g}_l
-\frac{{\bf g}_k^H{\bf \Pi}^{\bot}_{{\bf V}}{\bf u}_n{\bf u}^H_n{\bf \Pi}^{\bot}_{{\bf V}}{\bf g}_l}{{\bf u}^H_n{\bf \Pi}^{\bot}_{{\bf V}}{\bf u}_n}.
\end{equation}
Let us proceed now to  determine the expression of ${\bf g}_k$. Letting ${\bf A}'_{\theta_k} \pardef\frac{\partial {\bf A}_{\theta}}{\partial \theta_k}$, we get

\begin{equation}
\label{eq:Derivative Sigma}
\frac{\partial {\bf \Sigma}}{\partial \theta_k}={\bf A}'_{\theta_k}{\bf R}_s{\bf A}^H_{\theta}+{\bf A}_{\theta}{\bf R}_s{\bf A}^{'H}_{\theta_k},
\end{equation}
Hence, using \eqref{eq:ABC}, the \emph{kth} column of ${\bf G}$ in \eqref{eq:CRBGen} is given by
\begin{equation}
\label{eq:KelemG}
{\bf g}_k={\bf T}^{1/2}_i\ve({\bf Z}_k+{\bf Z}^H_{k})\ \ \mbox{where}\ \ {\bf Z}_k\pardef{\bf \Sigma}^{-1/2}{\bf A}_{\theta}{\bf R}_s{\bf A}^{'H}_{\theta_k}{\bf \Sigma}^{-1/2}.
\end{equation}
Next, we determine ${\bf V}$ and then ${\bf \Pi}^{\bot}_{{\bf V}}$.  Since ${\bf R}_{s}$ is a Hermitian matrix,  it can be then  factorized as
\begin{equation}
\label{eq:Vec.Rs}
\ve({\bf R}_s)={\bf J}\boldsymbol{\rho}
\end{equation}
where ${\bf J}$ is  a  $K^2\times K^2$  constant  nonsingular  matrix.  It follows, using \eqref{eq:ABC}, that ${\bf V}$  can be be expressed as
\[{\bf V}=  {\bf T}^{1/2}_i({\bf \Sigma}^{-T/2}{\bf A}^{*}_{\theta}\otimes{\bf \Sigma}^{-1/2}{\bf A}_{\theta}){\bf J}\pardef {\bf T}^{1/2}_i{\bf W}{\bf J}.\]
%
%
Note from \eqref{eq:CRBGen} that the SCRB depends on ${\bf V}$ only  via ${\bf \Pi}^{\bot}_{{\bf V}}$, that can be expressed as
\begin{equation}
\label{eq:Orthproj}
{\bf \Pi}^{\bot}_{{\bf V}}={\bf I}-{\bf V}({\bf V}^H{\bf V})^{-1}{\bf V}^H={\bf I}-{\bf T}^{1/2}_i{\bf W}({\bf W}^H{\bf T}_i{\bf W})^{-1}{\bf W}^H{\bf T}^{1/2}_i.
\end{equation}
After some algebraic manducation, using \eqref{eq:ABC} and \eqref{eq:ABCD}, we obtain
\[{\bf W}^H{\bf T}_i{\bf W}={\xi_2}({\bf U}^{*}\otimes{\bf U})+(\xi_2-1)\ve({\bf U})\ve^H({\bf U}),\]
where  ${\bf U}\pardef{\bf A}^{H}_{\theta}{\bf \Sigma}^{-1}{\bf A}_{\theta}$ is a $K\times K$  Hermitian nonsingular matrix. It follows from matrix inverse lemma (given by \eqref{eq:inv A B C D}), that its inverse can be expressed as
\[
({\bf W}^H{\bf T}_i{\bf W})^{-1}
=\frac{1}{\xi_2}({\bf U}^{-*}\otimes{\bf U}^{-1})-\eta\ve({\bf U}^{-1})\ve^H({\bf U}^{-1})
\]
where $\eta\pardef\frac{\xi_2-1}{\xi^2_2(1+\frac{\xi_2-1}{\xi_2}\ve^H(\tilde{\bf U})(\tilde{\bf U}^{-*}\otimes\tilde{\bf U}^{-1}) \ve(\tilde{\bf U}))}$ can be simplified, using \eqref{eq:trABCD}, as $\eta\pardef\frac{\xi_2-1}{\xi^2_2(1+\frac{\xi_2-1}{\xi_2}K)}$. Thus, using \eqref{eq:ABC} and  \eqref{eq:ABCD}, we obtain
\begin{equation}
\label{eq:MatB}
{\bf W}({\bf W}^H{\bf T}_i{\bf W})^{-1}{\bf W}^H
= \frac{1}{\xi_2}({\bf H}^{*}_1\otimes{\bf H}_1)-\eta\ve({\bf H}_1)\ve^H({\bf H}_1)\pardef{\bf \mathcal{B}},
\end{equation}
where ${\bf H}_1\pardef {{\bf \Sigma}}^{-1/2}{\bf A}_{\theta}{\bf U}^{-1}{\bf A}^H_{\theta}{{\bf \Sigma}}^{-1/2}$. Therefore,  \eqref{eq:Orthproj} becomes
\begin{equation}
\label{eq:Orthprojb}
{\bf \Pi}^{\bot}_{{\bf V}}={\bf I}-{\bf T}_i^{1/2}{\bf \mathcal{B}}{\bf T}_i^{1/2}.
\end{equation}

Now let us show that ${\bf u}^H_n{\bf \Pi}^{\bot}_{{\bf V}}{\bf g}_k=0$.  It follows from \eqref{eq:vecun} and \eqref{eq:KelemG}, using \eqref{eq:Orthprojb}, that
\begin{equation}
\label{eq:first.prod1}
{\bf u}^H_n{\bf \Pi}^{\bot}_{{\bf V}}{\bf g}_k
=
\ve^H({\bf \Sigma}^{-1}){\bf T}_i\ve({\bf Z}_k+{\bf Z}^H_{k})
-
\ve^H({\bf \Sigma}^{-1}){\bf T}_i{\bf \mathcal{B}}{\bf T}_i\ve({\bf Z}_k+{\bf Z}^H_{k}).
\end{equation}
It follows, after some algebraic manipulation, using \eqref{eq:ABC}, \eqref{eq:AB} and \eqref{eq:MatB} that 
\begin{eqnarray}
\nonumber
{\bf T}_i{\bf \mathcal{B}}{\bf T}_i&=&\xi_2({\bf H}^{*}_1\otimes{\bf H}_1)-\xi^2_2\eta\ve({\bf H}_1)\ve^H({\bf H}_1)
\\&+&
\nonumber
(\xi_2-1)(1-K\eta\xi_2)\left(\ve({\bf I})\ve^H({\bf H}_1)+\ve({\bf H}_1)\ve^T({\bf I})\right)
\\
&+&
\label{eq:TiBTi}
\frac{(\xi_2-1)^2K}{\xi_2}(1-K\eta\xi_2) \ve({\bf I})\ve^T({\bf I}),
\end{eqnarray}
using ${\bf H}^2_1={\bf H}_1$ and $\tra({\bf H}_1)=K$.
Using the definition \eqref{eq:def Ti} for ${\bf T}_i$ and \eqref{eq:AB}, the first term of \eqref{eq:first.prod1} can be expressed as
\begin{eqnarray}
\nonumber
\ve^H({\bf \Sigma}^{-1}){\bf T}_i\ve({\bf Z}_k+{\bf Z}^H_{k})
\!\!\!\!\!&=&\!\!\!\!\!
\xi_2\tra({\bf \Sigma}^{-1}({\bf Z}_k+{\bf Z}^H_{k}))+
(\xi_2-1)\tra({\bf \Sigma}^{-1})\tra({\bf Z}_k+{\bf Z}^H_{k})
\\
\label{eq:first.term1}
\!\!\!\!\!&=&\!\!\!\!\!
2\xi_2{\rm Re}(\tra({\bf \Sigma}^{-2}{\bf A}_{\theta}{\bf R}_s{\bf A}^{'H}_{\theta_k}))
\!+\!2(\xi_2\!-\!1)\tra({\bf \Sigma}^{-1}\!){\rm Re}(\tra({\bf \Sigma}^{-1}{\bf A}_{\theta}{\bf R}_s{\bf A}^{'H}_{\theta_k}))
\end{eqnarray}
using $\tra({\bf \Sigma}^{-1}({\bf Z}_k+{\bf Z}^H_{k}))=2{\rm Re}(\tra({\bf \Sigma}^{-2}{\bf A}_{\theta}{\bf R}_s{\bf A}^{'H}_{\theta_k}))$ and $\tra({\bf Z}_k+{\bf Z}^H_{k})=2{\rm Re}(\tra({\bf \Sigma}^{-1}{\bf A}_{\theta}{\bf R}_s{\bf A}^{'H}_{\theta_k}))$.
After simple algebraic manipulations, using \eqref{eq:TiBTi}, \eqref{eq:ABC} and \eqref{eq:AB}, and that $\tra({\bf Z}_k+{\bf Z}^H_{k})=\tra(({\bf Z}_k+{\bf Z}^H_{k}){\bf H}_1)=\tra({\bf H}_1({\bf Z}_k+{\bf Z}^H_{k}){\bf H}_1)=2{\rm Re}(\tra({\bf \Sigma}^{-1}{\bf A}_{\theta}{\bf R}_s{\bf A}^{'H}_{\theta_k}))$ and $\tra({\bf \Sigma}^{-1}{\bf H}^2_1)=\tra({\bf \Sigma}^{-1}{\bf H}_1)$, the second term of \eqref{eq:first.prod1} can be simplified as
\begin{eqnarray}
\nonumber
\ve^H({\bf \Sigma}^{-1}){\bf T}_i{\bf \mathcal{B}}{\bf T}_i\ve({\bf Z}_k+{\bf Z}^H_{k})
&&
\\
\nonumber
&&
\hspace{-3cm}
= \xi_2\tra({\bf \Sigma}^{-1}{\bf H}_1({\bf Z}_k+{\bf Z}^H_{k}){\bf H}_1)+(\xi_2-1)\tra({\bf \Sigma}^{-1})\tra({\bf Z}_k+{\bf Z}^H_{k})
\\
\nonumber
&&
\hspace{-3cm}
= 
2\xi_2{\rm Re}(\tra({\bf \Sigma}^{-1}{\bf A}_{\theta}{\bf U}^{-1}{\bf A}^H_{\theta}{\bf \Sigma}^{-2}{\bf A}_{\theta}{\bf R}_s{\bf A}^{'H}_{\theta_k}))
+
2(\xi_2-1)\tra({\bf \Sigma}^{-1}){\rm Re}(\tra({\bf \Sigma}^{-1}{\bf A}_{\theta}{\bf R}_s{\bf A}^{'H}_{\theta_k}))
\\
\label{eq:first.term2b}
&&
\hspace{-3cm}
=
2\xi_2{\rm Re}(\tra({\bf \Sigma}^{-2}{\bf A}_{\theta}{\bf R}_s{\bf A}^{'H}_{\theta_k}))+2(\xi_2-1)\tra({\bf \Sigma}^{-1}){\rm Re}(\tra({\bf \Sigma}^{-1}{\bf A}_{\theta}{\bf R}_s{\bf A}^{'H}_{\theta_k})),
\end{eqnarray}
where the first term in the last line  is obtained using ${\bf A}_{\theta}{\bf U}^{-1}{\bf A}^H_{\theta}{\bf \Sigma}^{-2}{\bf A}_{\theta}={\bf \Sigma}^{-1}{\bf A}_{\theta}$.
It follows, therefore, from \eqref{eq:first.prod1}, \eqref{eq:first.term1} and \eqref{eq:first.term2b} that 
$${\bf u}^H_n{\bf \Pi}^{\bot}_{{\bf V}}{\bf g}_k=0.$$
This identity together with \eqref{eq:KelemG} and \eqref{eq:Orthprojb}  allows us to rewrite the individual elements of \eqref{eq:CRBGen} as
\begin{eqnarray}
\nonumber
\frac{1}{T}\left[{\rm SCRB}^{-1}_{\rm CES}({\boldsymbol \theta})\right]_{k,l}
&=&
{\bf g}_k^H{\bf \Pi}^{\bot}_{{\bf V}}{\bf g}_l
\\
\label{eq:CRBGeneq1}
&=&
\ve^H({\bf Z}_k+{\bf Z}^H_k){\bf T}_i\ve({\bf Z}_l+{\bf Z}^H_l)
-\ve^H({\bf Z}_k+{\bf Z}^H_k){\bf T}_i{\bf \mathcal{B}}{\bf T}_i\ve({\bf Z}_l+{\bf Z}^H_l).
\end{eqnarray}
After simple algebraic manipulations, using the definition \eqref{eq:def Ti} for ${\bf T}_i$, \eqref{eq:ABC} and \eqref{eq:AB}, the first term in   \eqref{eq:CRBGeneq1} can be simplified as
\begin{eqnarray}
\nonumber
\ve^H({\bf Z}_k+{\bf Z}^H_k){\bf T}_i\ve({\bf Z}_l+{\bf Z}^H_l)
&=&
\xi_2\tra(({\bf Z}_k+{\bf Z}^H_{k})({\bf Z}_l+{\bf Z}^H_{l}))+
(\xi_2-1)\tra({\bf Z}_k+{\bf Z}^H_{k})\tra({\bf Z}_l+{\bf Z}^H_{l})\\&=& \nonumber2\xi_2\left[{\rm Re}(\tra(({\bf \Sigma}^{-1}{\bf A}_{\theta}{\bf R}_s{\bf A}^{'H}_{\theta_l})({\bf \Sigma}^{-1}{\bf A}_{\theta}{\bf R}_s{\bf A}^{'H}_{\theta_k})))\right. \\&+& \left.
\nonumber{\rm Re}(\tra(({\bf \Sigma}^{-1}{\bf A}^{'}_{\theta_l}{\bf R}_s{\bf A}_{\theta})({\bf \Sigma}^{-1}{\bf R}_s{\bf A}^{'H}_{\theta_k})))\right]
\\&+&\label{eq:Exp1}4(\xi_2-1){\rm Re}(\tra({\bf \Sigma}^{-1}{\bf A}_{\theta}{\bf R}_s{\bf A}^{'H}_{\theta_k})){\rm Re}(\tra({\bf \Sigma}^{-1}{\bf A}_{\theta}{\bf R}_s{\bf A}^{'H}_{\theta_l}))
\end{eqnarray}
Similarly, after some algebraic manipulations, using \eqref{eq:TiBTi}, \eqref{eq:ABC} and \eqref{eq:trABCD},  the second term in \eqref{eq:CRBGeneq1} can be simplified as
\begin{eqnarray}
\nonumber
\ve^H({\bf Z}_k+{\bf Z}^H_k){\bf T}_i{\bf \mathcal{B}}{\bf T}_i\ve({\bf Z}_l+{\bf Z}^H_l)&=& 2\xi_2\left[\tra({\rm Re}(({\bf \Sigma}^{-1}{\bf A}_{\theta}{\bf R}_s{\bf A}^{'H}_{\theta_l})({\bf \Sigma}^{-1}{\bf A}_{\theta}{\bf R}_s{\bf A}^{'H}_{\theta_k})))\right. \\&+& \left.
\nonumber\tra({\rm Re}(({\bf \Sigma}^{-1}{\bf A}{\bf U}^{-1}{\bf A}^H{\bf \Sigma}^{-1}{\bf A}^{'}_{\theta_l}{\bf R}_s{\bf A}^H_{\theta})({\bf \Sigma}^{-1}{\bf A}_{\theta}{\bf R}_s{\bf A}^{'H}_{\theta_k})))\right]
\\&+&\label{eq:Exp2}4(\xi_2-1)\tra({\rm Re}({\bf \Sigma}^{-1}{\bf A}_{\theta}{\bf R}_s{\bf A}^{'H}_{\theta_k}))\tra({\rm Re}({\bf \Sigma}^{-1}{\bf A}_{\theta}{\bf R}_s{\bf A}^{'H}_{\theta_l})).
\end{eqnarray}
It follows then from \eqref{eq:Exp1} and \eqref{eq:Exp2} that \eqref{eq:CRBGeneq1} can be simplified as
\begin{eqnarray}
 \nonumber
\frac{1}{T}\left[{\rm SCRB}^{-1}_{\rm CES}({\boldsymbol \theta})\right]_{k,l}&=&
2\xi_2{\rm Re}\left(\tra\left[({\bf \Sigma}^{-1}- {\bf \Sigma}^{-1}{\bf A}{\bf U}^{-1}{\bf A}^H{\bf \Sigma}^{-1})({\bf A}^{'}_{\theta_l}{\bf R}_s{\bf A}^{H}_{\theta}{\bf \Sigma}^{-1}{\bf A}_{\theta}{\bf R}_s{\bf A}^{'H}_{\theta_k})\right]\right)\\&=& \nonumber
\frac{2\xi_2}{\sigma^2_n}{\rm Re}\left(\tra\left[({\bf \Pi}^{\bot}_{{\bf A}_{\theta}})({\bf A}^{'}_{\theta_l}{\bf R}_s{\bf A}^{H}_{\theta}{\bf \Sigma}^{-1}{\bf A}_{\theta}{\bf R}_s{\bf A}^{'H}_{\theta_k})\right]\right)\\&=&\label{eq:CRBGeneq3}
\frac{2\xi_2}{\sigma^2_n}{\rm Re}\left(\tra\left[{\bf \Pi}^{\bot}_{{\bf A}_{\theta}}{\bf A}^{'}_{\theta_l}{\bf H}{\bf A}^{'H}_{\theta_k}\right]\right),
\end{eqnarray}
where the second equality is obtained using ${\bf \Sigma}^{-1}- {\bf \Sigma}^{-1}{\bf A}{\bf U}^{-1}{\bf A}^H{\bf \Sigma}^{-1}=\frac{1}{\sigma^2_n}{\bf \Pi}^{\bot}_{{\bf A}}$ thanks to ${\bf A}{\bf U}^{-1}{\bf A}^H{\bf \Sigma}^{-1}={\bf A}({\bf A}^H{\bf A})^{-1}{\bf A}^H$.
Using \eqref{eq:trABCD}, we can write \eqref{eq:CRBGeneq3} in matrix form as is shown in Result 5.

In the noncircular case, the proof follows the similar above steps     by replacing ${\bf T}_i$  by $\widetilde{\bf T}_i\pardef\frac{\xi_{2}}{2} {\bf I}+\frac{\xi_{2}-1}{4}\ve({\bf I})\ve^T({\bf I})$, and ${\bf \Sigma}$ by  $\widetilde{\bf \Gamma}$  where \eqref{eq:Derivative Sigma} is replaced by $\frac{\partial \widetilde{\bf \Gamma}}{\partial \theta_k}=\tilde{\bf A}'_{\theta_k}{\bf R}_{\tilde{s}}\tilde{\bf A}^H_{\theta}+\tilde{\bf A}_{\theta}{\bf R}_{\tilde{s}}\tilde{\bf A}^{'H}_{\theta_k}$ with $\tilde{\bf A}_{\theta}\pardef\diag({\bf A}_{\theta},{\bf A}^*_{\theta})$ and  $\tilde{\bf A}'_{\theta_k}\pardef\frac{\partial \tilde{\bf A}_{\theta}}{\partial \theta_k}$.
%
\section{Proof of Result 6}
\label{sec:Proof of Result 6}
The proof of this result  follows similar steps as the proof of Result 5  based on \cite[th. 1]{Abeida2017}  by replacing ${\bf \Sigma}$ by $\widetilde{\bf \Gamma}=\tilde{\bf A}_{\omega}{\bf R}_r\tilde{\bf A}^H_{\omega}+\sigma^2_n{\bf I}$, ${\bf A}_{\theta}$ by
$\widetilde{\bf A}_{\omega}
=
\left(\begin{array}{cc}
{\bf A}_{\theta}{\bf \Delta}_{\phi}\\
{\bf A}^*_{\theta}{\bf \Delta}^*_{\phi}\\
\end{array}
\right)$
where $\boldsymbol{\omega}\pardef (\boldsymbol{\theta}^T,\boldsymbol{\phi}^T)^T$
with $\boldsymbol{\phi} \pardef (\phi_1,...,\phi_K)^T$, and also by pointing out that   ${\bf R}_r\in\mathbb{R}^{K\times K}$ is symmetric which lead us to replace ${\bf J}$ in \eqref{eq:Vec.Rs} by ${\bf D}_{\rho}$ defined in \cite[th. 1]{Abeida2017} to get $\ve({\bf R}_r)={\bf D}_{\rho}\boldsymbol{\rho}$.  Thus, ${\bf V}$ becomes ${\bf V}=\widetilde{\bf T}^{1/2}_i{\bf W}{\bf D}_{\rho}$ with ${\bf W}=({\widetilde{\bf \Gamma}}^{-T/2}{\widetilde{\bf A}}^{*}_{\omega}\otimes{\widetilde{\bf \Gamma}}^{-1/2}\widetilde{\bf A}_{\omega})$.  Hence ${\bf \Pi}^{\bot}_{{\bf V}}$ in \cite[th. 1]{Abeida2017} takes here the following key form expression:  
${\bf \Pi}^{\bot}_{{\bf V}}={\bf I}-\widetilde{\bf T}^{1/2}_i{\bf \mathcal{B}}\widetilde{\bf T}^{1/2}_i$ with ${\bf \mathcal{B}}
=\frac{2}{\xi_2} {\bf W}({\bf U}^{-1}\otimes{\bf U}^{-1}){\bf N}_K{\bf W}^H-\tilde{\eta}\ve({\bf H}_1)\ve^H({\bf H}_1)$ 
where ${\bf U}\pardef\widetilde{\bf A}^{H}_{\omega}\widetilde{\bf \Gamma}^{-1}\widetilde{\bf A}_{\omega}$, ${\bf N}_{K}$ is defined in \cite[th. 1]{Abeida2017} and $\tilde\eta\pardef\frac{\xi_2-1}{\xi^2_2(1+\frac{\xi_2-1}{2\xi_2}K)}$.
The rest of the proof follows the same lines of arguments as that of the proof of  Result 5.

\end{document}